\documentclass[%
superscriptaddress,
twocolumn,
amsmath,amssymb,
 aps,
prb,
floatfix,
]{revtex4-2}

\usepackage{siunitx}
\usepackage{graphicx}
\usepackage{csquotes}
\usepackage[colorlinks]{hyperref}
\hypersetup{%
	plainpages=true,
	breaklinks=true,
	hypertexnames=false,
	pageanchor=true,
	colorlinks=true,
	linkcolor={blue},
	citecolor={red},
	urlcolor={blue},
	anchorcolor={black}
}

\usepackage[dvipsnames]{xcolor}

\begin{document}
\title{First- and second-order quantum phase transitions in the long-range unfrustrated antiferromagnetic Ising chain}

\author{Víctor Herráiz-López}
\thanks{These two authors contributed equally}
\affiliation {Departamento de Física Aplicada, Universidad de Zaragoza, Zaragoza 50009, Spain}

\author{Sebastián Roca-Jerat}
\thanks{These two authors contributed equally}
\affiliation {Instituto de Nanociencia y Materiales de Aragón (INMA), CSIC-Universidad de Zaragoza, Zaragoza 50009, Spain}
\affiliation{Departamento de Física de la Materia Condensada, Universidad de Zaragoza, Zaragoza 50009, Spain}

\author{Manuel Gallego}
\affiliation {Instituto de Nanociencia y Materiales de Aragón (INMA), CSIC-Universidad de Zaragoza, Zaragoza 50009, Spain}
\affiliation{Departamento de Física Teórica, Universidad de Zaragoza, Zaragoza 50009, Spain}

\author{Ramón Ferrández}
\affiliation {Instituto de Nanociencia y Materiales de Aragón (INMA), CSIC-Universidad de Zaragoza, Zaragoza 50009, Spain}
\affiliation{Departamento de Física de la Materia Condensada, Universidad de Zaragoza, Zaragoza 50009, Spain}

\author{Jesús Carrete}
\affiliation {Instituto de Nanociencia y Materiales de Aragón (INMA), CSIC-Universidad de Zaragoza, Zaragoza 50009, Spain}
\affiliation{Departamento de Física de la Materia Condensada, Universidad de Zaragoza, Zaragoza 50009, Spain}

\author{David Zueco}
\affiliation {Instituto de Nanociencia y Materiales de Aragón (INMA), CSIC-Universidad de Zaragoza, Zaragoza 50009, Spain}
\affiliation{Departamento de Física de la Materia Condensada, Universidad de Zaragoza, Zaragoza 50009, Spain}

\author{Juan Román-Roche}
\affiliation {Instituto de Nanociencia y Materiales de Aragón (INMA), CSIC-Universidad de Zaragoza, Zaragoza 50009, Spain}
\affiliation{Departamento de Física de la Materia Condensada, Universidad de Zaragoza, Zaragoza 50009, Spain}

\date{\today}

\begin{abstract}
    We study the ground-state phase diagram of an unfrustrated antiferromagnetic Ising chain with longitudinal and transverse fields in the full range of interactions: from all-to-all to nearest-neighbors. First, we solve the model analytically in the strong long-range regime, confirming in the process that a mean-field treatment is exact for this model. We compute the order parameter and the correlations and show that the model exhibits a tricritical point where the phase transition changes from first to second order. This is in contrast with the nearest-neighbor limit where the phase transition is known to be second order. To understand how the order of the phase transition changes from one limit to the other, we tackle the analytically-intractable interaction ranges numerically, using a variational quantum Monte Carlo method with a neural-network-based ansatz, the visual transformer. We show how the first-order phase transition shrinks with decreasing interaction range and establish approximate boundaries in the interaction range for which the first-order phase transition is present. Finally, we establish that the key ingredient to stabilize a first-order phase transition and a tricritical point is the presence of ferromagnetic interactions between spins of the same sublattice on top of antiferromagnetic interactions between spins of different sublattices. Tunable-range unfrustrated antiferromagnetic interactions are just one way to implement such staggered interactions.
\end{abstract}

\maketitle

\section{Introduction}
\label{sec:intro}

Recent advances in cold-atom simulators have led to renewed interest in systems with long-range interactions \cite{Britton2012, knap2013probing, monroe2021programmable, browaeys2020manybody, scholl2021quantum}. Long-range interactions decay as $r^{-\alpha}$, with $r$ the distance between interacting degrees of freedom \cite{mukamel2008statistical, campa2009statistical, defenu2023longrange, defenu2023outofequilibrium}. Despite being ubiquitous in nature, {e.g.} dipolar or gravitational interactions, these systems have been less studied than their nearest-neighbor counterparts because long-range interactions complicate numerical and analytical treatments. Nevertheless, some results exist that showcase remarkable differences in behavior between long- and short-range models. Some examples are the spontaneous breaking of continuous symmetries in one dimension \cite{maghrebi2017continuous}, the existence of an area law of entanglement \cite{koffel2012entanglement, kuwahara2020area, vodola2014kitaev, ares2018entanglement}, the existence of Majorana modes \cite{jager2020edge}, the spreading of correlations \cite{schneider2022spreading} and topological properties \cite{viyuela2016topological, gong2016topological}. 
Notably, long-range interactions have been shown to induce first-order phase transitions in classical dipolar gases \cite{cartarius2014structural} and (quantum) Bose-Hubbard \cite{landig2016quantum, blab2018quantum} and XX \cite{igloi2018quantum} models.

Quantum phase transitions constitute one of the fundamental phenomena of condensed matter physics. They manifest as nonanaliticities in the ground state energy as some critical parameter is varied \cite{sachdev2011}. Leaving aside the more exotic topological phase transitions \cite{berezinsky1970destruction, kosterlitz1972long, kosterlitz1973ordering}, quantum phase transitions, like thermal phase transitions, can be first or second order, depending on whether the order parameter or its derivatives are discontinuous at the critical point.
Just as the Ising model is the paradigmatic model in classical statistical mechanics, its quantum analogue, the (nearest-neighbour) transverse field Ising chain (TFIC), is the paradigmatic example of a solvable model featuring a quantum phase transition \cite{biganmbeng2024the}. The TFIC is also a starting point to devise more sophisticated models.
%
\begin{figure}[b]
    \centering
    \includegraphics[width =0.37\columnwidth]{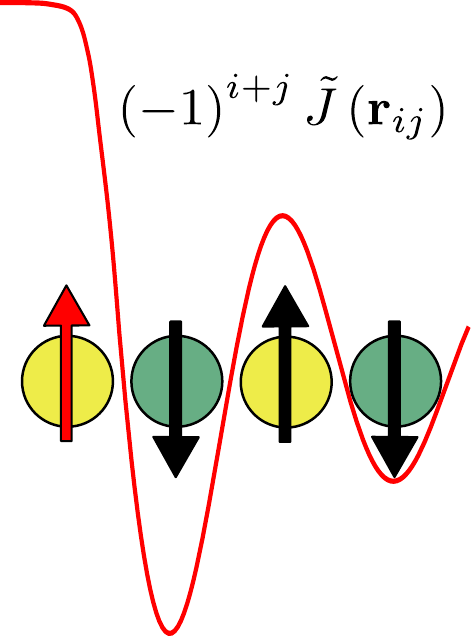}
    \caption{Sketch of long-range unfrustrated antiferromagnetic interactions between a given spin (in red) and its first three neighboring spins. The interactions are ferromagnetic between spins belonging to the same sublattice and antiferromagnetic between spins belonging to different sublattices.}
    \label{fig:sketch_interactions}
\end{figure}
To elucidate the effect of long-range interactions on the order of phase transitions,  it is convenient to construct a minimal model. The trivial long-range generalization of the TFIC is known to feature only a second-order phase transition \cite{koziol2021quantumcritical, sun2017fidelity, rocajerat2024transformer}. In fact, there is no phase transition at all for antiferromagnetic interactions at extremely long ranges. At the same time, it has been shown that the Ising model with antiferromagnetic nearest-neighbor interactions and ferromagnetic next-nearest-neighbor interactions presents a tricritical point (TP) where the critical line changes from first to second order \cite{kato2015quantum}.
To shed more light on this issue, we have generalized the antiferromagnetic Ising chain to feature tunable-range \emph{unfrustrated} antiferromagnetic interactions and studied its ground-state phase diagram analytically in the strong long-range regime using the technique described in Ref.~\citenum{romanroche2023exact}. We find that the phase diagrams of the nearest-neighbor and strong long-range models differ significantly. The nearest-neighbour model presents a second-order critical line between an antiferromagnetic and a paramagnetic phase, with a singular point -for vanishing transverse field, where the model becomes classical- where the phase transition becomes first order \cite{novotny1986zero, sen2000quantum, ovchinnikov2003antiferromagnetic, simon2011quantum, neto2013phase}. In contrast, our results show that in the strong long-range model, a significant portion of the critical line between the antiferromagnetic and paramagnetic phases becomes first order. The critical line changes order at a tricritical point that occurs at non-zero transverse field. 
To understand how the critical line morphs from one limit to the other, we employ a variational quantum Monte Carlo method with a neural-network-based ansatz, the visual transformer \cite{dosovitskiy2021image}, in the full range of interactions $0 < \alpha < \infty$, with $\alpha \to \infty$ corresponding to the nearest-neighbor limit. These numerical results show how the position of the tricritical point smoothly moves from zero to a finite transverse field. Remarkably, the first-order phase transition survives well into the weak long-range regime ($\alpha > d$ with $d$ the dimension of the lattice, and specifically $d=1$ for the chain). Finally, we check whether the first-order phase transition is present if the ferromagnetic intrasublattice interactions are removed. In this case, we find that the full critical line is second order. This indicates that the key to stabilize a first-order phase transition is the simultaneous presence of antiferromagnetic intersublattice and ferromagnetic intrasublattice interactions. Tunable-range antiferromagnetic interactions are just one way to implement and tune these staggered interactions.

The rest of the paper is organized as follows. In Sec. \ref{sec:model} we present the Hamiltonian for the tunable-range unfrustrated antiferromagnetic Ising chain. Section \ref{sec:method} is dedicated to the exact solution of the model in the strong long-range regime, with a characterization of the ground state phase diagram through the order parameter and the correlations. The numerical results are presented in Sec. \ref{sec:numerical}. We end the paper with a discussion of the results and the source of the first-order phase transition in Sec. \ref{sec:conclusions} and provide technical details and complementary results in the appendices.

\section{The model}
\label{sec:model}

We consider a one-dimensional spin chain ($d=1$) with tunable-range Ising interactions and subject to transverse and longitudinal fields. The Hamiltonian reads
\begin{equation}
    H = -\omega_z \sum_{i=1}^N S_i^z - \omega_x \sum_{i=1}^N S_i^x - \sum_{i,j=1}^N J_{ij}S_i^xS_j^x \; ,
    \label{eq:Hamiltonian}
\end{equation}
where $S_i^{x,z}$ are spin-$s$ operators acting on site $i$. The interactions are $J_{ij} = \left(-1\right)^{i+j} \Gamma \tilde J(\mathbf r_{ij})/\tilde N$, with
\begin{equation}
	\tilde J(\mathbf r_{ij})= \begin{cases}
	b & \text { if } \quad \mathbf r_{ij}=0 \\
	|\mathbf r_{i j}|^{-\alpha} & \text { otherwise}
	\end{cases}
 \label{eq:J_tilde}
\end{equation}
where the distance $\mathbf{r}_{ij}$ is given by the nearest image convention using periodic boundary conditions (PBC). $\Gamma >0$ is the interaction strength, $b$ is a parameter that can be tuned to shift the spectrum of $J$, $\tilde N = \sum_i \tilde J_{ij}$ is Kac's renormalization factor and $\alpha$ is the coefficient that tunes the range of interaction. The alternating sign makes the interactions antiferromagnetic when spins are separated by an odd number of lattice parameters and ferromagnetic when spins are separated by an even number of parameters, see Fig. \ref{fig:sketch_interactions}. According to the classification of long-range interactions valid for both classical and quantum models, the strong long-range regime corresponds to $\alpha < d=1$ \cite{mukamel2008statistical,defenu2023longrange}. This regime is characterized by the loss of additivity, although extensivity is preserved by Kac's renormalization factor, $\tilde N$, ensuring a well-defined thermodynamic limit. In the limit $\alpha \to \infty$, the Hamiltonian [Eq.~\eqref{eq:Hamiltonian}] corresponds to the nearest-neighbor Ising chain with transverse and longitudinal fields.
The tunable-range staggered interactions that we consider here are a generalization of the antiferromagnetic nearest-neighbor interactions in unfrustrated lattices. The alternating sign prevents frustration and allows for the formation of two sublattices, as sketched in Fig.~\ref{fig:sketch_interactions}. For vanishing fields, the ground state is the antiferromagnetic configuration, with the spins fully polarized along the $x$ axis in alternating directions for even and odd spins. Staggered tunable-range interactions have been studied previously for the Heisenberg model \cite{yusuf2004spin, laflorencie2005critical, beach2007valence, sandvik2010ground, gong2016topological, luhang2021from, ren2022longrange, adelhart2023continuously, zhao2024unconventional, adelhardt2024quantumcritical}.

\section{Exact solution in the strong long-range regime}
\label{sec:method}

\subsection{Canonical partition function}
\label{sec:Z}
Reference~\citenum{romanroche2023exact} presents an exact analytical solution for quantum strong long-range models in the canonical ensemble and the thermodynamic limit ($N \to \infty$), extending previous classical results \cite{campa2003canonical}. We sketch its application to Hamiltonian \eqref{eq:Hamiltonian} here; the details can be found in App.~\ref{app:phi_antiferro}.

The Hamiltonian is divided into free and interacting parts
\begin{equation}
    H = H_0 - \sum_{i,j}J_{ij}S_i^x S_j^x \; ,
\end{equation} 
First, the interaction matrix is diagonalized
\begin{equation}
    J_{ij} = \frac{1}{N} \sum_{k=0}^{N-1} \lambda_{ik} D_k \lambda_{jk} \; ,
    \label{eq:interaction_matrix}
\end{equation}
where $D_k$ are its eigenvalues and $\lambda_{ik}/\sqrt{N}$ is an orthogonal matrix because $J_{ij}$ is symmetric. The smallest eigenvalue is set to zero by tuning the on-site interaction parameter $b$ \eqref{eq:J_tilde}. For $s\neq 1/2$, setting $b\neq0$ introduces new terms in the Hamiltonian, but due to Kac's renormalization factor, $\tilde N$, these are negligible in the thermodynamic limit.
Then, the Hamiltonian can be mapped to a generalized Dicke model \cite{dicke1954coherence}, replacing the long-range interactions by effective interactions between each particle and a set of $M$ real fields, $\left\{ u_k \right\}$, where $M$ is the number of non-zero eigenvalues of $J$.
The mapping is only valid if $\lim_{N\to\infty} M/N = 0$, restricting the applicability of the method to models in which the number of non-zero eigenvalues of the interaction matrix is a negligible fraction of the total.
This is the reason why the solution is restricted to the strong long-range regime, where $\alpha<d$ ensures that this condition is met.
Following Wang's procedure, we can obtain the canonical partition function of the corresponding Dicke model \cite{wang1973phase,hioe1973phase}, thus solving our original model. This yields
\begin{equation}
    Z = A \exp \left(-N \beta f[\bar u_k]\right)
    \label{eq:Z_saddle_point}
\end{equation}
where $f[\bar u_k] = \min_{\left\{u_k\right\}} f [u_k]$, with $f [u_k] = f(u_0, \ldots, u_{M-1})$ and $\beta = 1 / k_{\rm B} T$, with $k_{\rm B}$ Boltzman's constant and $T$ the temperature, and
\begin{equation}
    A = \left({\left(\frac{\beta}{2}\right)^M \, {\det} H_f [\bar u_k] }\right)^{-1/2} \prod_{p=0}^{M-1} \sqrt{\frac{D_p}{\beta}} \; ,
\end{equation}
with $H_f [\bar u_k]$ the Hessian of $f[u_k]$ at its global minimum.
By noticing that in the thermodynamic limit $f[\bar u_k]$ is the free energy per particle: $f[\bar u_k] = \lim_{N \to \infty} - (N\beta)^{-1} \ln Z$, it becomes apparent that computing the partition function is reduced to a minimization of the variational free energy per particle
\begin{equation}
    f[u_k] = \sum_{k=0}^{M-1} \frac{u_k^2}{D_k} + f_{\rm m}[u_k] 
    \label{eq:fvariational}
\end{equation}
with respect to the auxiliary fields $\{u_k\}$. Here $f_{\rm m} = F_{\rm m} / N$, with
\begin{equation}
    F_{\rm m}\left[u_k\right] = -\frac{1}{\beta} \sum_{i=1}^N \ln \left[  \frac{\sinh \left( (2s+1)\beta \varepsilon_i[u_k] \right)}{\sinh \left(\beta \varepsilon_i[u_k] \right)}  \right] \;,
    \label{eq:F_m}
\end{equation}
and
\begin{equation}
     2\varepsilon_i [u_k] = \sqrt{\omega_z^2 + \left( \omega_x + 2\sum_{k=0}^{M-1} \lambda_{ik} u_k  \right)^2} \; .
     \label{eq:varepsilon}
\end{equation}

Although computing the canonical partition function has been reduced to solving a minimization problem, finding the global minima of a multivariate function is not a simple task, and success is not guaranteed in most cases. However, for $\alpha = 0$ (homogeneous all-to-all couplings which are ferromagnetic or antiferromagnetic depending on the distance between the spins) the problem naturally becomes univariate, as there is only one non-zero eigenvalue of the interaction matrix $J$, so $M=1$. This is equivalent to setting $u_{k \neq 0} = 0$, with $u_0$ corresponding to the largest eigenvalue: $D_0 = \max \{D_k\} = \Gamma$ and $\lambda_{i0} = (-1)^i$. Additionally, we find that this solution with $u_{k \neq 0} = 0$ is the global minimum also for $0 \neq \alpha < 1$. This can be shown analytically for $\omega_x = 0$ (See. Appendix \ref{app:omega_x_0}) and has been verified numerically otherwise. With this, computing the canonical partition function has been reduced to a univariate minimization of $f(u) \equiv f(u, 0, \ldots, 0)$ in terms of $u \equiv u_0$ for all $\alpha < 1$, which can be tackled analytically or numerically. In the remainder of this section, we do so to obtain the ground state phase diagram of the model and the susceptibilities. This also implies that the equilibrium properties of the model are universal for all $\alpha < 1$. In fact, $f(u)$ coincides with the free energy obtained with a mean-field approach, proving that mean field is exact also for strong long-range unfrustrated antiferromagnetic models. This is consistent with the fact that unfrustrated antiferromagnetic models can be mapped to ferromagnetic models in a staggered field and a mean-field description of strong long-range ferromagnetic models has been shown to be exact \cite{mori2012equilibrium}.

\subsection{Ground State Phase Diagram}
\label{sec:phase_diagram}

\begin{figure}
    \centering
    \includegraphics[width = \columnwidth]{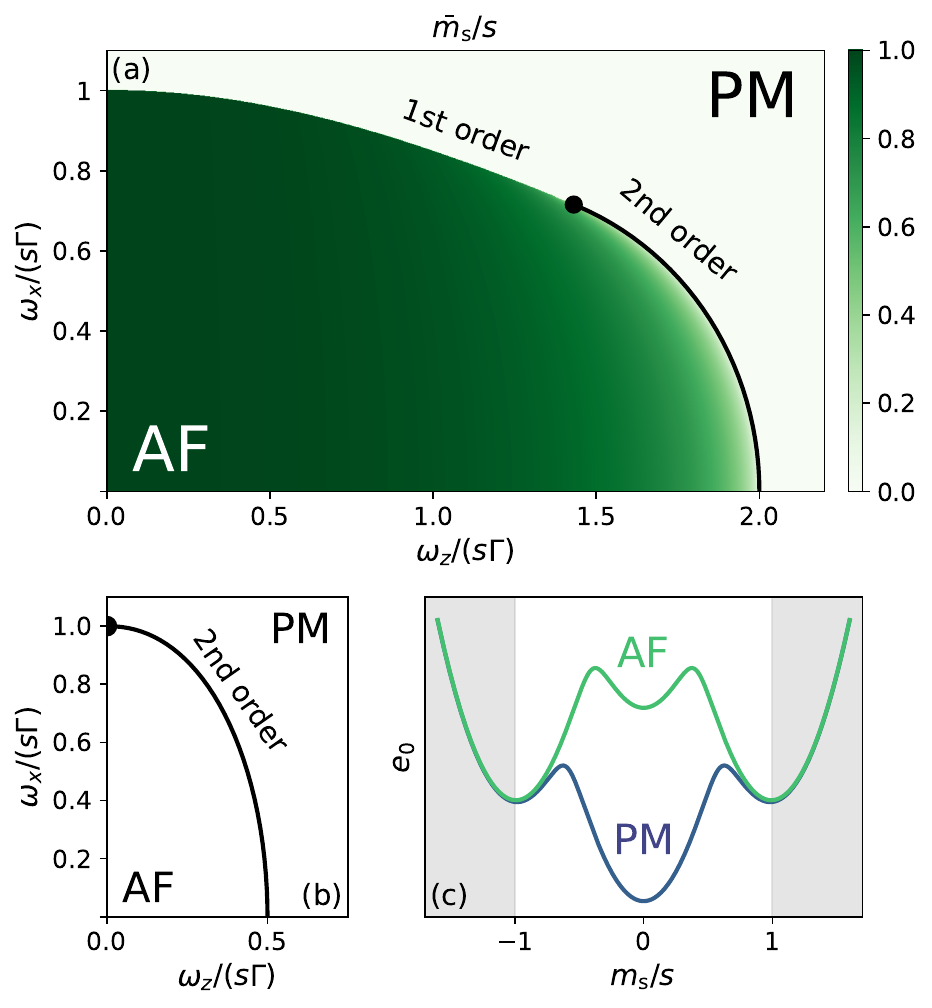}
    \caption{(a) Ground-state phase diagram of Hamiltonian \eqref{eq:Hamiltonian} for $\alpha < 1$ as a function of the longitudinal, $\omega_x$, and transverse, $\omega_z$, fields. 
    The black line marks the second-order portion of the critical line. The black dot marks the tricritical point. (b) Sketch of the ground-state phase diagram of Hamiltonian \eqref{eq:Hamiltonian} for $\alpha=\infty$. (c) Variational free energy \eqref{eq:variational_energy} in the antiferromagnetic, $(\omega_x,\omega_z)=(0.8,0.2)s\Gamma$, and paramagnetic, $(\omega_x,\omega_z)=(1.2,0.2)s\Gamma$, phases. Grey shaded areas are not valid values of the staggered magnetization, but are included for improved visualization of the energy landscape.}
    \label{fig:multiplot_phase_diagram}
\end{figure}

In order to study the ground state of the model (See App. \ref{app:nonzerotemp} for the finite temperature phase diagram), we define the variational ground-state energy as the zero-temperature limit of the variational free energy $e_0(u) = \lim_{\beta\to\infty} f(u)$, such that
\begin{equation}
    e_0(u) = \frac{u^2}{\Gamma}
    - s \left( \varepsilon_+(u) +  \varepsilon_-(u)\right) \,,
\label{eq:variational_energy} 
\end{equation}
with
\begin{equation}
    2 \varepsilon_\pm(u) = \sqrt{\omega_z^2 + (\omega_x \pm 2u)^2} \,.
    \label{eq:varepsilonunivariate}
\end{equation}
It can be shown that the global minimum, $\bar u$, is proportional to the staggered magnetization (See App. \ref{app:magnetization_susceptibilities}) 
\begin{equation}
    \bar m_{\rm s} = \frac{1}{N} \sum_{i=1}^N \left(-1\right)^i \langle S_i^x \rangle = \frac{\bar u}{\Gamma} \; . \label{eq:staggered_magnetization}
\end{equation}
The staggered magnetization is the order parameter of unfrustrated antiferromagnetic models such as Hamiltonian \eqref{eq:Hamiltonian}. It is zero in the paramagnetic phase and it measures how close the ground state is to a perfect antiferromagnetic configuration in the antiferromagnetic phase. Figure \ref{fig:multiplot_phase_diagram}(a) shows the phase diagram of the model. The staggered magnetization is obtained from $\bar u$, by minimizing $e_0$, and plotted as a function of the longitudinal and transverse fields. The model exhibits a quantum phase transition (QPT) between an antiferromagnetic phase and a paramagnetic phase. The nature of the phase transition changes along the critical line from a second- to a first-order QPT. The change can be described analytically by applying the Landau theory of phase transitions to a series expansion of $e_0(m_{\rm s})$, with $m_{\rm s} = u/\Gamma$ the variational staggered magnetization.  By studying the change of sign of the expansion coefficients we obtain the tricritical point
\begin{align}
    \omega_{x, \rm tp} = s\Gamma \frac{8}{5\sqrt{5}} \; , \label{eq:tricritical_x} \\
     \omega_{z, \rm tp} = s\Gamma \frac{16}{5\sqrt{5}} \; , \label{eq:tricritical_z}
\end{align}
and the equation for the second-order portion of the critical line
\begin{equation}
    4s^2\Gamma^2 \omega_z^4 = \left( \omega_z^2 + \omega_x^2 \right)^3 \; ,
\end{equation}
when $\vert \omega_z \vert > 2\vert \omega_x \vert$.
The first-order critical condition cannot be obtained from this analysis because the global minima corresponding to the antiferromagnetic configurations fall outside of the radius of analyticity of the series expansion of $e_0(m_{\rm s})$. Nevertheless, having identified the second-order portion, the rest of the critical line is first order by exclusion. We verify this graphically in Fig. \ref{fig:multiplot_phase_diagram}(c) where we show the landscape of minima of $e_0$ for different values of $\omega_x$ when $\omega_z / (s\Gamma) = 0.2$. For $\omega_x / (s\Gamma)=0.8$ there exists a local minimum at $m_{\rm s}=0$, corresponding to the paramagnetic state, and two degenerate global minima at $m_{\rm s} \approx \pm 1$, corresponding to the symmetric antiferromagnetic ground states. For $\omega_x / (s \Gamma)=1.2$, the minimum at $m_{\rm s} = 0$ has become the global minimum,  following a first-order phase transition between the antiferromagnetic and paramagnetic phases. The same behavior is observed for any $\omega_z < \omega_{z, \rm tp}$.

We have thus found the existence of a first-order QPT on a finite portion of the critical line. We emphasize that this is valid for all strong long-range models, $\alpha < 1$, and that this behaviour is still present at finite temperatures (See App. \ref{app:nonzerotemp}). This constitutes a qualitative difference with respect to the nearest-neighbor limit, $\alpha \to \infty$, of the model, in which the QPT is second-order along the full critical line except for the point where $\omega_z = 0$, when the model is classical \cite{simon2011quantum,ovchinnikov2003antiferromagnetic}. This is sketched in Fig. \ref{fig:multiplot_phase_diagram}(b) for comparison.  It is worth noting that there is some evidence that the phase transition could be mixed order \cite{bar2014mixed, mukamel2023mixed} around the classical point in the nearest-neighbor limit, but not purely first order like we have just described for the strong long-range regime \cite{lajko2021mixedorder}.  

%
%

For $\alpha = 0$ it is possible to do a classical analysis of the model that predicts the same phase diagram that we just described, with distinct finite regions of first- and second-order phase transitions. See App. \ref{app:classical} for more details. This illustrates that a classical analysis can serve as an exploratory tool for spin models with all-to-all interactions. It offers an alternative perspective on the phase transition based on the magnetizations of each sublattice rather than the order parameter (the staggered magnetization in this case). In any case, it is not a replacement for an exact solution due to the fact that its application is limited to $\alpha = 0$ and that it does not allow for the computation of susceptibilities. 

\subsection{Analysis of correlations}

Introducing a perturbative field to the Hamiltonian~\eqref{eq:Hamiltonian}, $H \to H -\sum_{i=1}^N h_i S_i^x$, we can compute the susceptibilities as
\begin{equation}
    \chi_{ij} = \frac{1}{\beta}\lim_{\{h_n\}\to 0} \frac{\partial^2 \ln Z[h_n]}{\partial h_j \partial h_i}  \; . \label{eq:suscept_main_text}
\end{equation}

The susceptibility is proportional to the Kubo correlator \cite{kubo1991statistical}[Chap. 4], and hence a measure of correlations between spins.
For a translation invariant model the susceptibility can be computed analytically \cite{schwabl}[Chap. 6] (see details in App. \ref{app:magnetization_susceptibilities}). At zero temperature it can be written as
\begin{equation}
    \chi_{ij} = (A^{-1})_{ij} Y_j \,,
    \label{eq:susclosedexp}
\end{equation}
where the matrix is defined by $A_{ij} = \delta_{ij} - 2Y_iJ_{ij}$ and $Y_i = s \omega_z^2 / (8 \bar \varepsilon_i)$ with $\bar \varepsilon_i = \varepsilon_{(-1)^i}(\bar u)$ as defined in Eq.~\eqref{eq:varepsilonunivariate}.

\begin{figure}
    \centering
    \includegraphics[width = \columnwidth]{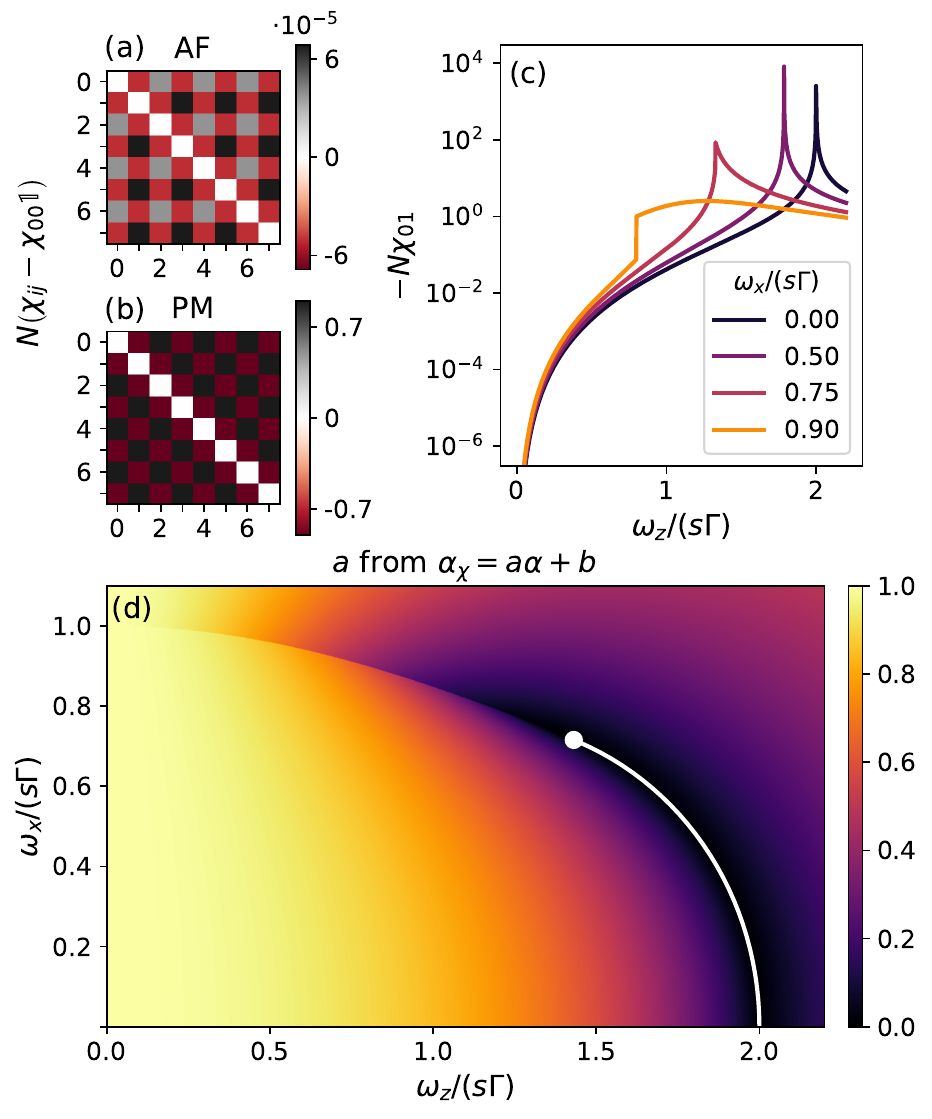}
    \caption{(a) Susceptibility matrix for $\alpha=0$ in the antiferromagnetic phase, $(\omega_x,\omega_z)=(0.1,0.2)s\Gamma$. (b) Susceptibility matrix for $\alpha=0$ in the paramagnetic phase, $(\omega_x,\omega_z)=(1,2)s\Gamma$. Both have been obtained for $N=8$ and autocorrelations have not been depicted. (c) Evolution of intersublattice correlations across the phase diagram, computed using $N=100$ and $\alpha=0$. (d) Slope of the linear relation $\alpha_\chi=a\alpha+b$, computed using $N=100$. It is independent of the element from the susceptibility matrix used. The white line marks the second-order portion of the critical line. The white dot marks the tricritical point.}
    \label{fig:multiplot_correlations}
\end{figure}

In Figs.~\ref{fig:multiplot_correlations}(a) and (b) we show the behavior of the susceptibility matrix, $\chi_{ij}$, for $\alpha=0$ in the two phases of the model. Although Eq. \eqref{eq:susclosedexp} has been obtained in the thermodynamic limit ($N \to \infty$), evaluating it requires that we fix a finite value of $N$. We set $N=8$ and mask out the diagonal elements to better display the structure of the correlation matrix. For $\alpha = 0$, in the antiferromagnetic phase the correlations show an alternating pattern with intrasublattice correlations being positive and intersublattice correlations being negative. In addition, the intrasublattice correlations of the sublattice that is aligned with the longitudinal field (even sites) are weaker than the intrasublattice correlations of the sublattice that is aligned against the field (odd sites). This effect accentuates with increasing longitudinal field, to the point that the correlations almost vanish for the sublattice aligned with the field for large longitudinal field. The spins of the sublattice that is aligned with the longitudinal field behave increasingly as paramagnetic free spins despite the model still being in an antiferromagnetic state. In the paramagnetic phase the two sublattices become equally magnetized along the combined external field and the correlation matrix recovers a perfectly alternating pattern that matches the staggered interactions, with intra- and intersublattice correlations being of equal magnitude and opposite sign.

Setting $\alpha \neq 0$ introduces a spatial dependence in the correlations in both phases. The structure of the correlation matrix remains the same but the elements are modulated by the distance between the corresponding spins. 
The correlations exhibit a power-law decay with distance within each of the possible families: intersublattice, even intrasublattice and odd intrasublattice. This is the case regardless of the proximity to the critical point. This behaviour is typical of strong long-range systems and contrasts with weak long-range and short range systems where power-law decay is only present at the critical point \cite{vodola2015longrange, vanderstraeten2018quasiparticles, francica2022correlations, romanroche2023exact}. Since the model is translation invariant, we can define $\chi_{r, 01} \equiv \chi_{0 \,2r+1}$, $\chi_{r, 00} \equiv \chi_{0 \, 2r}$ and $\chi_{r, 11} \equiv \chi_{1 \, 2r+1}$ for the intersublattice, even intrasublattice and odd intrasublattice correlations; they all follow a power-law decay: $\chi_r \propto r^{-\alpha_\chi}$ with the same exponent, $\alpha_\chi$. 
In the paramagnetic phase 
$\chi_{r, 00} = \chi_{r, 11}$. The rate of decay of correlations depends linearly on the rate of decay of interactions, i.e. $\alpha_\chi = a \alpha + b$. Fig.~\ref{fig:multiplot_correlations} (d) shows the slope, $a$, across the phase diagram. The numerical fit of the relation also shows that $b \approx 0$ in all cases, which is consistent with the fact that the model must become distance independent for $\alpha = 0$. Close to the second-order critical line, $\alpha_\chi$ becomes independent of $\alpha$, with $a = 0$. This is consistent with the behavior described in Ref. \citenum{romanroche2023exact} for the strong long-range ferromagnetic Ising model. In contrast, the first-order phase transition is marked by a discontinuity in the slope of the linear dependence between two non-zero values.

The effect of the first- and second-order phase transitions is also apparent in the behavior of single correlation matrix elements across the phase diagram. Figure \ref{fig:multiplot_correlations}(c) shows the value of a correlation matrix element as a function of the longitudinal and transverse fields, for $\alpha = 0$. They exhibit a divergence at the second-order phase transition and a finite discontinuity at the first-order phase transition. The behavior is analogous for any matrix element and any value of $\alpha$. 

%
%

\section{Numerical solution for arbitrary interaction range}
\label{sec:numerical}

Having established a significant difference between the phase diagram of the strong long-range and nearest-neighbors models, it is natural to wonder how this difference is interpolated for arbitrary values of the range of interactions, $\alpha$. Since only the strong long-range regime is analytically tractable, in this section we resort to zero-temperature variational quantum Monte Carlo (qMC) \cite{newman1999monte} simulations with a visual transformer (ViT) \cite{dosovitskiy2021image, viteritti2023transformer, viteritti2024spinliquid} ansatz for finite system sizes to study the model in the full range of interactions. We show the results obtained for spins with $s = 1/2$ using this architecture, given its recent success in describing spin-1/2 chains with long-range interactions \cite{rocajerat2024transformer}. Details on the particular implementation used here can be found in App.~\ref{app:vithyperparams}.

Using this technique we obtain the ground state of finite-size chains with periodic boundary conditions along the whole parameter space ($\omega_x$, $\omega_z$, $\alpha$). In order to study how the phase transition evolves with $\alpha$, we choose as order parameter the squared staggered magnetization $m_s^2$ [cf. Eq.~\eqref{eq:staggered_magnetization}]. This choice is justified by the fact that, in finite-size chains, the symmetry of the ground state is not broken and the staggered magnetization is zero throughout the whole parameter space. The squared magnetization, on the contrary, exhibits all the features related to first- and second-order phase transitions and allows us to carry out this numerical characterization.

\begin{figure}
    \centering
    \includegraphics[width=\columnwidth]{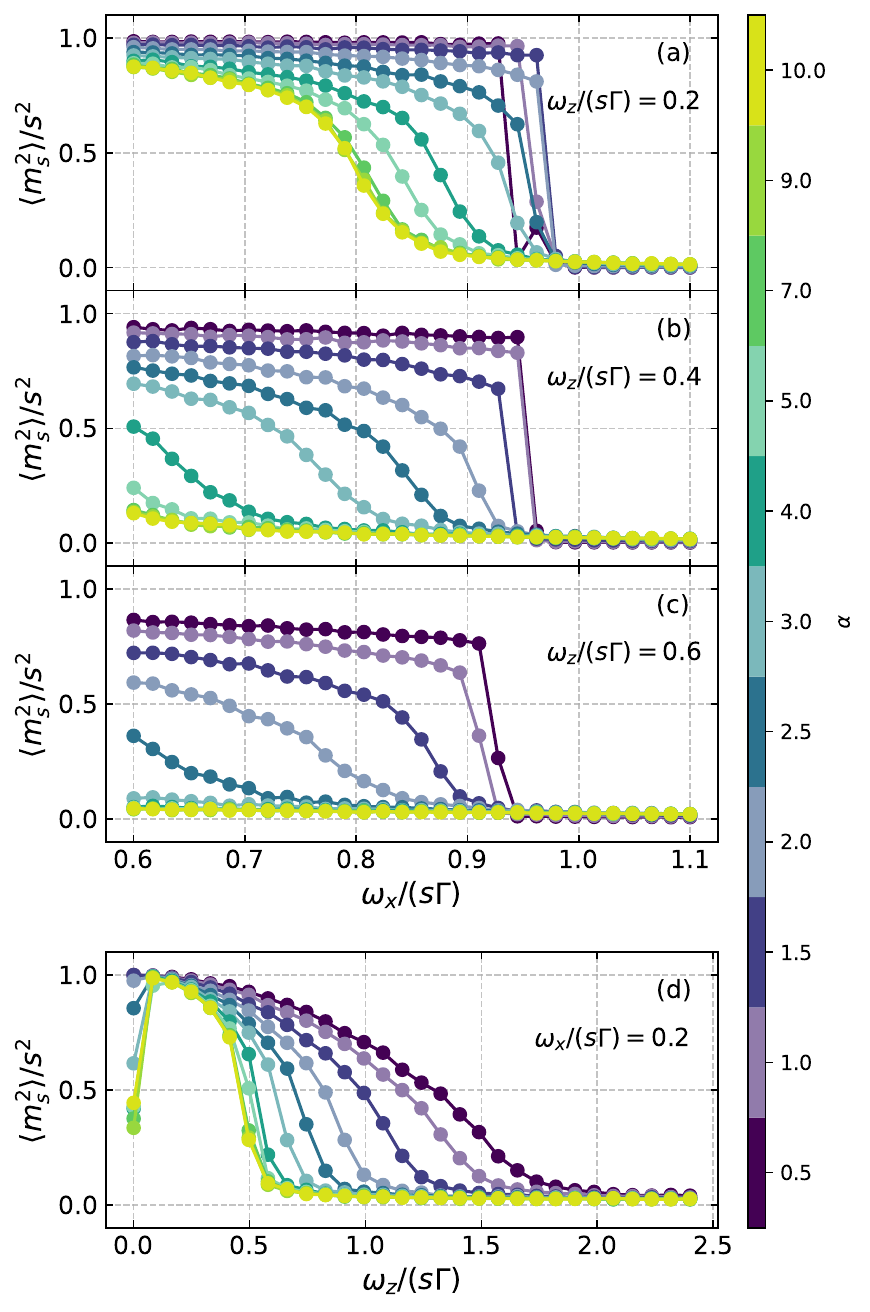}
    \caption{Characterization of the parameter space through vertical slices (panels a,b,c) and horizontal slices (panel d) of the phase diagram for the squared staggered magnetization. Each slice represents a different $\alpha$ value and all simulations have been carried out for a chain size of $N = 50$.}
    \label{fig:phasediagramcuts_alpha}
\end{figure}

In Fig. \ref{fig:phasediagramcuts_alpha} we observe how $\langle m_s^2\rangle$ behaves for different slices of the phase diagram ({cf.} Fig.~\ref{fig:multiplot_phase_diagram}) as a function of the different values of $\alpha$ for a fixed chain size of $N = 50$. In panels (a), (b) and (c), which correspond to vertical slices (fixed transverse field), we see that for certain values of $\alpha$ the nature of the phase transition evolves from first to second order as the transverse field, $\omega_z$, increases. Panel (a), corresponding to $\omega_z/(s\Gamma) = 0.2$, merits special attention. It shows that the discontinuity characteristic of first-order phase transitions is clearly present up to $\alpha = 2.5$ (See App.~\ref{app:morenumericalresults} for further confirmation). In panels (b) and (c) we see how this transition becomes second order, with a continuous change of the order parameter at the critical point. If instead of comparing a given value of $\alpha$ across panels we focus on any given panel, (a), (b) or (c), we observe that the transition develops a discontinuity as $\alpha$ is lowered. In all cases the transition is clearly discontinuous for $\alpha \leq 1$, in agreement with the analytical predictions. 

It is interesting to note that analytical results of Sec. \ref{sec:method} predict that for any $\alpha \leq 1$, the critical behavior should be universal. In contrast, the numerical results do exhibit a dependence of the critical point and the value of the order parameter on $\alpha$. We attribute this to finite-size effects, which are expected to be particularly important in long-range systems. The relatively small size considered here allows us to witness dependencies on $\alpha$ that are washed away in the thermodynamic limit. Additionally, we observe that the ansatz encounters difficulties in obtaining the correct ground state near the critical point for $\alpha \leq 1$. One would expect the critical point to recede toward smaller values of $\omega_x$ as $\alpha$ increases. However, for $\alpha \leq 1$ the states corresponding to the ordered and paramagnetic phases are practically degenerate in energy and the ViT fails to systematically converge to the correct ground state, so this trend can no longer be numerically confirmed.

In Fig. \ref{fig:phasediagramcuts_alpha}(d) we show a horizontal slice of the phase diagram (fixed longitudinal field). The phase transition is second order for all values of $\alpha$, as expected from the analytical results of Sec.~\ref{sec:method}. Following Figs.~\ref{fig:multiplot_phase_diagram}(a) and (b) we expected a decrease of the second-order critical point as $\alpha$ increases. This is confirmed by the numerical results.
The critical point tends to the predicted values of $\omega_z/(s\Gamma) = 2$ for $\alpha \leq 1$ (strong long-range regime) and $\omega_z/(s\Gamma) = 0.5$ in the limit of $\alpha \to \infty$ (nearest-neighbors regime). We observe that from $\alpha \gtrsim 7$ the curves collapse and thus it can be considered as the numerical limit from which the model operates in the nearest-neighbors regime. On the other hand, the numerical instabilities now appear at a point where the model is classical (no transverse field).

\begin{figure}
    \centering
    \includegraphics[width=\columnwidth]{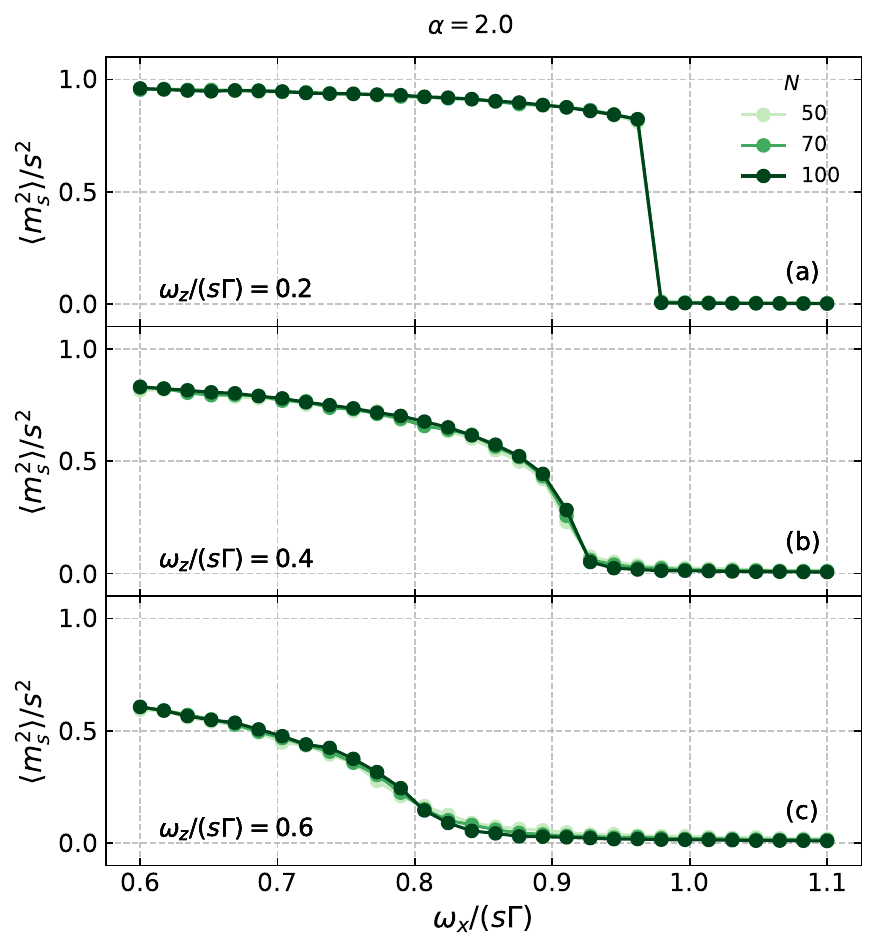}
    \caption{Squared staggered magnetization behavior for three different vertical slices of the phase diagram and $\alpha = 2$. The first-order transition transforms into a second-order phase transition as the transverse field is increased. Lighter colors indicate smaller sizes, being $N = 50$ the smallest, $N = 100$ the largest and $N = 70$ the intermediate size.}
    \label{fig:vertcut_alpha2p0}
\end{figure}

To better certify the order of the different phase transitions, beyond a visual analysis of the discontinuities (or lack thereof) of the order parameter, we perform a finite-size scaling analysis. In Fig. \ref{fig:vertcut_alpha2p0} we show this analysis for $\alpha = 2$ (See App.~\ref{app:morenumericalresults} for other values of $\alpha$). As it can be seen from panel (a), all the curves corresponding to $N = 50, 70, 100$ coincide, showing an independence with size characteristic of first-order transitions, in addition to the discontinuity in the order parameter. In contrast, in panel (c), the curves exhibit the typical size dependence of second-order phase transitions. The continuous change of the order parameter becomes increasingly non-analytical as the size increases. The crossover point marks the second-order critical point expected in the thermodynamic limit.

\begin{figure}
    \centering
    \includegraphics[width = \columnwidth]{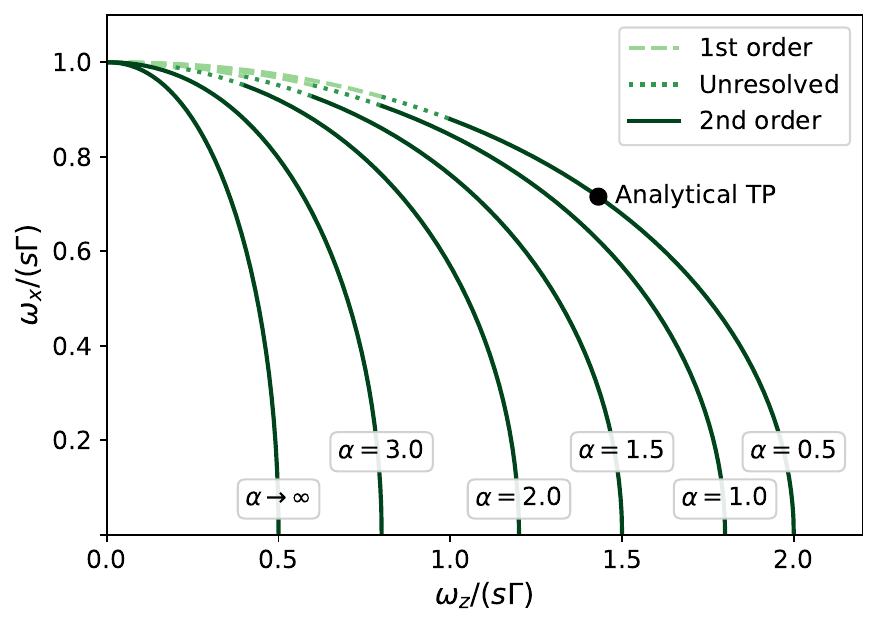}
    \caption{Summary of the numerical results for the order of the quantum phase transition. For reference, the black dot marks the analytical tricritical point in the strong long-range regime [Eqs.~\eqref{eq:tricritical_x} and \eqref{eq:tricritical_z}].}
    \label{fig:complete_sketch}
\end{figure}

Figure \ref{fig:complete_sketch} summarizes the numerical results discussed in this section, along with additional results reserved for App.~\ref{app:morenumericalresults}. We sketch the critical line for different values of $\alpha$, highlighting the regions where it is first and second order. We have established that the first-order phase transition at finite transverse field is present for $\alpha \leq 2.5$. The portion of the critical line that is first order decreases progressively with increasing $\alpha$ until for $\alpha \gtrsim 3$ the full critical line is second order. This is remarkable because it indicates that the first-order phase transition is present not only in the strong long-range regime but also beyond the mean-field threshold and presumably (see the next paragraph) in the full weak long-range regime. Although the precise boundaries, $\alpha_{\rm MF}$ and $\alpha^*$, between the regime in which the model exhibits mean-field critical exponents and the weak long-range regime and between the weak long-range and short-range regimes, have not been established for the antiferromagnetic model under consideration, it is known that for the analogous ferromagnetic model they lie at $\alpha_{\rm MF} = 5/3 \approx 1.66$ and $\alpha^* = 3$ \cite{defenu2023longrange}.
Additionally, we have witnessed the shrinking of the antiferromagnetic phase as the model approaches the short range regime, driven by a decrease of the critical transverse field at zero longitudinal field from $\omega_{z}/(s\Gamma) = 2$ for $\alpha > 1$ to $\omega_{z}/(s\Gamma) = 0.5$ for $\alpha \gtrsim 7$. 

It is worth noting that the first-order phase transition is very sensitive to finite-size effects. This is evident from the curve corresponding to $\alpha = 0.5$. We know from our analytical results that the tricritical point where the transition changes from first to second order occurs at $\omega_z / (s\Gamma) = 16/(5 \sqrt{5}) \approx 1.43$. However, from our numerical results with $N=100$ we are only able to certify a first-order phase transition up to $\omega_z / (s\Gamma) = 0.8$. This leads us to believe that our numerical results are also underestimating the region where the critical line is first order for all the other values of $\alpha$, and therefore also for how low a value of $\alpha$ the first-order phase transition survives.

\section{Discussion}
\label{sec:conclusions}

In this paper we have generalized the Ising chain in transverse field to feature tunable-range antiferromagnetic interactions. We have solved the model analytically in the strong long-range regime, showing that it presents a tricritical point in the phase diagram where the critical line changes from first to second order. This is in contrast with the nearest-neighbours limit, where the critical line was known to be always second order but for the point of vanishing transverse field, where the model is classical. To understand the transition from one limit to the other, we have studied the model numerically in the full range of interactions. We have found that the first-order phase transition is present beyond the strong long-range regime, apparently in the full weak long-range regime, only disappearing when the model enters the short-range regime beyond $\alpha \approx 3$. This confirms that the range of interactions can influence the nature of phase transitions in a model. 

The fact that similar behaviour has been observed in a model with only antiferromagnetic nearest-neighbor and ferromagnetic next-nearest-neighbor interactions \cite{kato2015quantum} suggests that the key ingredient that stabilizes a first-order phase transition is the presence of intrasublattice ferromagnetic interactions, in contrast with models with only intersublattice antiferromagnetic interactions. To verify this hypothesis we have also considered Hamiltonian \eqref{eq:Hamiltonian} but without intrasublattice ferromagnetic interactions, i.e. with $J_{ij} = 0$ for $i+j$ even. This model is not tractable with the analytical method used in Sec.~\ref{sec:method}, so an exact solution in the strong long-range regime is not possible. Nevertheless, a classical analysis of the $\alpha = 0$ limit, analogous to the one described in App.~\ref{app:classical}, shows that the phase transition is always second order, except for the point of vanishing transverse field, $\omega_z = 0$, where the model is classical. Thus, we can conclude that intrasublattice ferromagnetic interactions (on top of intersublattice antiferromagnetic interactions) are the key ingredient to generate a first-order phase transition. In that sense, in the tunable-range unfrustrated antiferromagnetic Ising model of Hamiltonian \eqref{eq:Hamiltonian}, $\alpha$ acts as a tuning knob to change the ratio between inter- and intrasublattice interactions. For low enough $\alpha$, intrasublattice ferromagnetic interactions become large enough to stabilize a first-order phase transition. The same mechanism might explain the first-order phase transitions reported in other quantum long-range models with staggered interactions \cite{landig2016quantum, blab2018quantum, igloi2018quantum}.

Finally, the appearance of a first-order phase transition at zero temperature suggests that the strong long-range unfrustrated antiferromagnetic Ising chain may exhibit ensemble inequivalence at finite temperature. Ensemble inequivalence has been reported to appear in the strong long-range regime in both classical \cite{hou2020from, barre2001inequivalence} and quantum models featuring a first-order phase transition \cite{russomanno2020quantum, defenu2024ensemble}.

\section*{Acknowledgements}
The authors acknowledge funding from the grant
TED2021-131447B-C21 funded by MCIN/AEI/10.13039/501100011033 and the EU
‘NextGenerationEU’/PRTR, grant CEX2023-001286-S financed by MICIU/AEI /10.13039/501100011033, the Gobierno de Aragón (Grant E09-17R Q-MAD), Quantum Spain and the CSIC Quantum
Technologies Platform PTI-001. This research project was made possible through the access granted by the Galician Supercomputing Center (CESGA) to its supercomputing infrastructure. The supercomputer FinisTerrae III and its permanent data storage system have been funded by the Spanish Ministry of Science and Innovation, the Galician Government and the European Regional Development Fund (ERDF). J. R-R acknowledges support from the Ministry of Universities of the Spanish Government through the grant FPU2020-07231. S. R-J.  acknowledges financial support from Gobierno de Arag\'on through a doctoral fellowship.
\appendix

\section{Solving the staggered antiferromagnetic Ising chain}
\label{app:phi_antiferro}

After diagonalizing the interaction matrix, $J$, as in Eq.~\eqref{eq:interaction_matrix}, Hamiltonian~\eqref{eq:Hamiltonian} can be mapped onto a generalized Dicke model \cite{dicke1954coherence, hioe1973phase} given by
\begin{equation}
    H_D = H_0 + \sum_{k=0}^{M-1} \frac{1}{D_k} a_k^\dagger a_k - \sum_{k=0}^{M-1} \sum_{i=1}^N \left(a_k + a_k^\dagger \right) \frac{\lambda_{ik}}{\sqrt{N}} S_i^x \; ,
\end{equation}
where $H_0 = -\omega_z\sum_i^N S_i^z - \omega_x \sum_i^N S_i^x$ and $a_k$ and $a_{k^\prime}^\dagger$, with $[a_k,a_{k^\prime}^\dagger] = \delta_{k,k^\prime}$, are the creation and annihilation operators of $M$ bosonic modes.
The mapping is valid if $\lim_{N\to\infty} M/N=0$, where $M$ is the number of non-zero eigenvalues of the interaction matrix, $J$, from the Hamiltonian~\eqref{eq:Hamiltonian} (the smallest eigenvalue is always set to zero by tuning the parameter $b$ from Eq.~\eqref{eq:J_tilde}).
We have thus replaced the interaction between the spins by an interaction between each spin and a set of $M$ bosonic modes.

Following Wang's procedure \cite{wang1973phase,hioe1973phase}, the canonical partition function of that generalized Dicke model can be obtained by first computing the partial trace over the matter degrees of freedom.
Then, replacing the photonic degrees of freedom by a collection of $M$ real Gaussian integrals, we obtain
\begin{equation}
    Z = \int \prod_{k=0}^{M-1}\sqrt{\frac{ND_k}{\pi\beta}}du_k \, \exp\left\{-N\beta f[u_k]\right\} \; ,
    \label{eq:Z_integral_appendix}
\end{equation}
where the variational free energy per particle, $f[u_k]$, is given by Eq.~\eqref{eq:fvariational}.

The variational free energy has two contributions: one accounting only for the real fields associated with the bosonic modes, $\left\{u_k\right\}$, and the function $F_{\rm m}[u_k]$ [Eq.~\eqref{eq:fvariational}] which accounts for the spins and their interaction with the bosonic modes. It is computed as
\begin{equation}
    F_{\rm m}[u_k] = -\frac{1}{\beta} \ln\left(Z_{\rm m}\right) \; ,
\end{equation}
with
\begin{equation}
        Z_{\rm m}[u_k] = {\rm Tr_m}\left[ \exp\left\{ -\beta \left( H_0 - \sum_{i=1}^N \sum_{k=0}^{M-1} 2\lambda_{ik} u_k S_i^x \right) \right\}  \right] \; . 
\end{equation}
$Z_{\rm m}[u_k]$ represents the partial trace of the generalized Dicke model over its matter degrees of freedom. Using the fact that it factorizes over the $N$ different spins we obtain
\begin{equation}
\begin{split}
    Z_{\rm m}[u_k] = \prod_{i=1}^N \left( \sum_{m=-s}^{+s} \exp\left\{-2m\beta\varepsilon{u_k}\right\} \right) \\
    = \prod_{i=1}^N \frac{\sinh\left((2s+1)\beta\varepsilon_i\right)}{\sinh\left(\beta\varepsilon_i\right)} \; ,
\end{split}
\end{equation}
where $\varepsilon_i[u_k]$ is given by Eq.\eqref{eq:varepsilon}.

Hence, using Wang's procedure the partition function of the original model can be expressed as the integral over a set of $M$ real fields, Eq.~\eqref{eq:Z_integral_appendix}.
The explicit linear dependence on $N$ of the exponent allows using the saddle-point method, which is exact in the limit $N\to\infty$.
Hence, the integral can be replaced by the value of its integrand at its maximum, as done in Eq.~\eqref{eq:Z_saddle_point}, which corresponds to finding the global minimum of the variational free energy per particle.

\section{Finding the global minimum of the variational free energy per particle in absence of longitudinal field}
\label{app:omega_x_0}

We want to find the global minimum of the variational free energy per particle, $f\left[ u_k \right]$ from Eq.~\eqref{eq:fvariational}, for all $\alpha < 1$ and $\omega_x = 0$.
Defining for each site on the lattice the variable
\begin{equation}
    \mu_i = \sum_{k=0}^{M-1} \lambda_{ik} u_k \; , \label{eq:app_mu}
\end{equation}
allows writing Eq,~\eqref{eq:varepsilon} as
\begin{equation}
    2\varepsilon_i[u_k] = \sqrt{\omega_z^2 + 4 \mu_i^2} \; . \label{eq:app_varepsilon_mu}
\end{equation}
The relation from Eq.~\eqref{eq:app_mu} can be inverted to obtain $u_k = \sum_{i=1}^N \mu_i \lambda_{ik}/N$.

The vanishing gradient condition of $f[u_k]$ in its local minima, $\left\{\bar u_k \right\}$, can be written as
\begin{equation}
    \frac{\bar u_k}{D_k} = \frac{s}{N} \sum_{i=1}^N \lambda_{ik} \frac{\sum_{l=0}^{M-1}\lambda_{il}\bar u_l}{\bar \varepsilon_i} B_s(2s\beta\bar \varepsilon_i ) \; ,
\end{equation}
with $B_s(x)$ the Brillouin function ($B_{1/2}(x) = \tanh(x)$) \cite{kittel2005introduction}[Chap. 11] and $\bar \varepsilon_i = \varepsilon_i[\bar u_k ]$.
A linear combination of this set of $M$ equations allows us to find a relation satisfied in the stationary points of the variational free energy,
\begin{equation}
    \bar \mu_j = s \sum_{i=1}^N J_{ij} \frac{\bar \mu_i}{\bar \varepsilon_i} B_s (2s\beta \bar \varepsilon_i) \; .
\end{equation}
This relation allows us to particularize the expression of $f[\bar u_k]$, {i.e.} the expression of the variational free energy per particle in its stationary points, as
\begin{equation}
\begin{split}
    &f[\bar u_k] \\
    &= \frac{1}{N} \sum_{j=1}^N \left[ s\beta\frac{\bar\mu_j^2}{\bar\varepsilon_j} B_s(2s\beta\bar\varepsilon_j) - \ln\left( \frac{\sinh((2s+1)\beta\bar\varepsilon_j)}{\sinh(\beta\bar\varepsilon_j)} \right)   \right] \; .
\end{split}
\end{equation}
Hence, in the stationary points the variational free energy can be written as a sum of a certain function in each site of the lattice
\begin{equation}
    f[\bar u_k] = \frac{1}{N}\sum_{i=1}^N \xi(\bar \mu_i^2) \; , \label{eq:f_sum_xi}
\end{equation}
as $\bar\varepsilon_i$ does not depend on $\bar \mu_i$ itself but on $\bar \mu_i^2$ (see Eq.~\eqref{eq:app_varepsilon_mu}).

Equation \eqref{eq:f_sum_xi} means that among all the possible local minima, the global one must satisfy that $\vert \bar \mu_i \vert = \bar \mu$ for each lattice site, in order to minimize each of the $N$ terms $\xi(\bar \mu_i^2)$.
Due to the structure of the matrix $\lambda_{ik}$, the only possible configuration that meets $\vert \bar \mu_i \vert = \bar \mu$ is
\begin{equation}
    \bar \mu_i = (-1)^i \bar \mu \; .
\end{equation}
It can be translated into a condition for the fields $\{u_k\}$ inverting the relation from Eq.~\eqref{eq:app_mu}. This condition must be satisfied in the global minimum of the variational free energy per particle,
\begin{align}
    \bar u_{k\neq0}=\bar \mu  \, \delta_{k0} \; .
\end{align}
Hence, it analytically shows that the multivariate minimization problem can be turned into a single-variable one when $\omega_x = 0$ for any value of $\alpha$.
Then, when $\alpha=0$ the problem is naturally single-variable as $M=1$ for any value of $\omega_x$.
For any other $\alpha<1$ it has been numerically checked that the problem remains single-variable, as was analytically proved in the strong long-range ferromagnetic model \cite{romanroche2023exact}.

\section{Analytical study of the non-zero temperature phase diagram}
\label{app:nonzerotemp}

In the strong long-range regime, we have reduced computing the canonical partition function to solving a minimization problem.
We have to find the global minimum of the variational free energy per particle, Eq.~\eqref{eq:fvariational}, with respect to the auxiliary fields $\left\{ u_k \right\}$.
We have discussed that the problem becomes univariate for every $\alpha < 1$, i.e. in the global minimum $\bar u_k = u_0 \delta_{k0}$.
Although in the main text we focus on the ground state phase diagram, we can also use this technique to study the finite temperature properties of the model \citenum{romanroche2023exact}.

In the ground state phase diagram we have found the existence of a critical line whose order shifts from first to second order at a tricritical point.
This is in contrast with the nearest-neighbors limit, where the critical line remains of second order except on the vanishing-transverse-field limit.
In this appendix we will show that this long-range behavior of the phase diagram is still present for finite temperatures.

\begin{figure}
    \centering
    \includegraphics[width = \columnwidth]{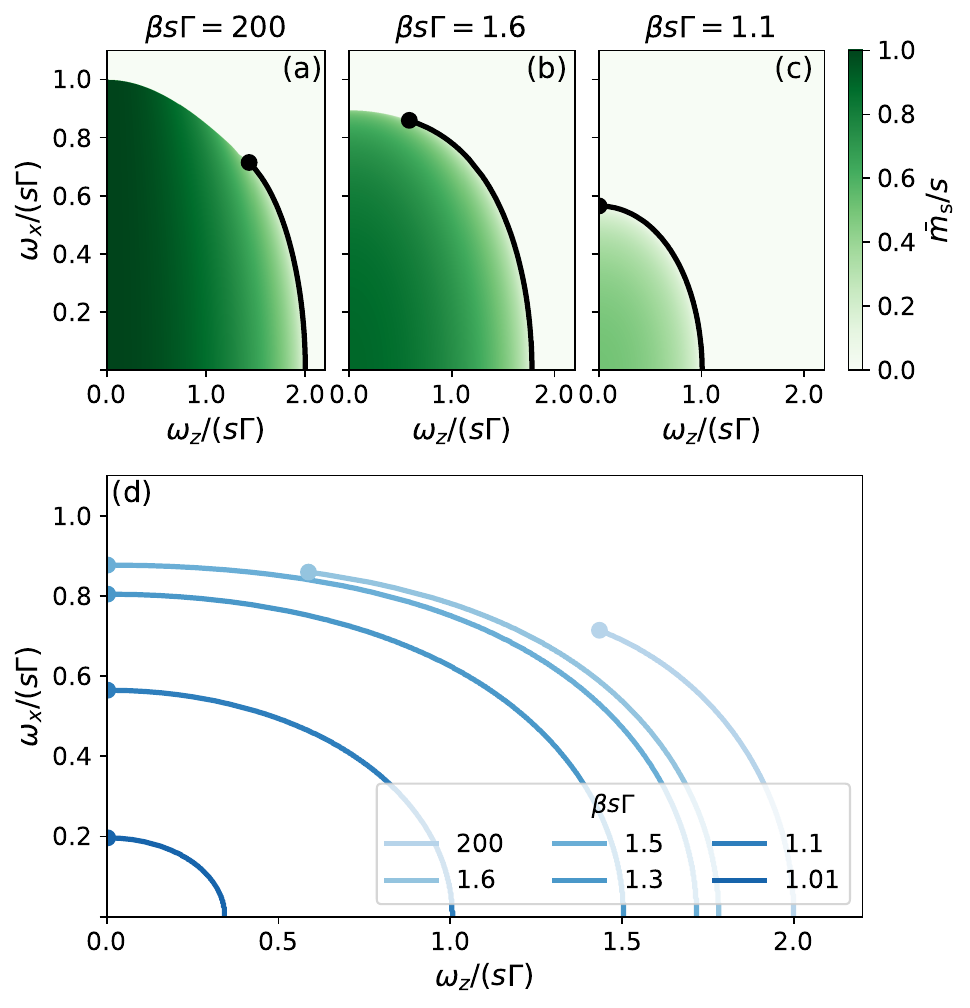}
    \caption{(a) Comparison between the phase diagram obtained via numerical minimization of the variational free energy and the second order critical line obtained using Landau theory, using $\beta s \Gamma = 200$. (b) $\beta s \Gamma = 1.6$. (c) $\beta s \Gamma = 1.1$. (d) Thermal evolution of the second-order critical line and the tricritical point.}
    \label{fig:multiplot_nonzerotemp}
\end{figure}

To study the phase diagram we numerically minimize the univariate free energy per particle, $f(u_0)$.
To find the second-order critical line and the tricritical point we also use Landau theory.
In Fig.~\ref{fig:multiplot_nonzerotemp}(a), (b) and (c) we show the phase diagram obtained via numerical minimization of the variational free energy and the second-order critical line obtained using Landau Theory.

In Fig.~\ref{fig:multiplot_nonzerotemp}(d) we summarize the evolution of the second-order critical line as temperature increases ($\beta s\Gamma$ decreases).
What we find is that increasing temperature reduces the ordered phase, and it also reduces the portion of the critical line corresponding to a first-order phase transition.
Up to the numerically achieved accuracy, the first-order critical line disappears for $\beta s\Gamma < 3/2$, and the ordered phase finally disappears for $\beta s\Gamma < 1$.

\section{Analytical calculation of the magnetization and the susceptibility}
\label{app:magnetization_susceptibilities}

As explained in the main text, introducing a perturbative logitudinal field in Hamiltonian~\eqref{eq:Hamiltonian} allows us to obtain the magnetization of the model and the susceptibilities between spins. In this appendix we will present the analytical exact expression for both magnitudes in the strong long-range regime.

By introducing a perturbative magnetic field, Hamiltonian~\eqref{eq:Hamiltonian} is replaced by $H \to H - \sum_{i=1}^N h_s S_i^x$, which leads to a partition function which depends on the perturbative field: $Z[h_n] = Z(h_1,\dots,h_N)$, defined by Eq. \eqref{eq:Z_saddle_point} with the replacement $\omega_x \to \omega_x + h_i$ in Eq.~\eqref{eq:varepsilon}.
To compute the perturbative-field-dependent partition function, the variational free energy will depend on $\{h_n\}$ through Eq.~\eqref{eq:varepsilon}, modifying the minimization problem we need to solve.

First, we can define field-dependent magnetization and susceptibilities as
\begin{align}
    \langle S_i^x \rangle [h_n] = \frac{1}{\beta} \frac{\partial \ln Z\left[h_n\right]}{\partial h_i} \; , \label{eq:magnetization_field_dependent} \\
    \chi_{ij}[h_n]  = \frac{1}{\beta} \frac{\partial^2 \ln Z\left[h_n\right]}{\partial h_j \partial h_i} \; , \label{eq:susceptibilities_field_dependent}
\end{align}
which allow us to compute the magnetization and the susceptibilities from Hamiltonian \eqref{eq:Hamiltonian} as $\langle S_i^x \rangle = \lim_{\{h_n\}\to 0} \langle S_i^x \rangle [h_n]$ and $\chi_{ij} = \lim_{\{h_n\}\to 0} \chi_{ij} [h_n]$.
For a translation invariant model, as considered, we will be able to compute both magnitudes analytically \cite{schwabl}[Chap. 6].

A direct computation of the derivative \eqref{eq:magnetization_field_dependent} allows us to write
\begin{equation}
    \langle S_i^x  \rangle \left[ h_n \right] = s B_s\left(2s\beta \bar \varepsilon_i \right) \frac{\omega_x + h_i + 2 \sum_{k=0}^{M-1} \Bar{u}_k \lambda_{ik}}{2 \bar \varepsilon_i} \;.
    \label{eq:expected_value_S_h}
\end{equation}
Note that here the solution to the minimization problem will depend on the perturbative fields, i.e. $\bar u_k \equiv \bar u_k[h_n]$.

Using Eq.~\eqref{eq:expected_value_S_h} the vanishing gradient condition of the variational free energy per particle in its global minimum can be writen as
\begin{equation}
   \frac{1}{D_k} \bar u_k = \frac{1}{N} \sum_{i=1}^N \lambda_{ik} \langle S_i^x \rangle [h_n] \; . \label{eq:app_null_gradient_condition}
\end{equation}
In the antiferromagnetic model $\lambda_{i0}=(-1)^i$ and $D_0=\Gamma$, which means that $\bar u_0/\Gamma = \bar m_{\rm s}$, proving Eq.~\eqref{eq:staggered_magnetization} in the limit $\{h_n\}\to 0$ as a consequence of the null gradient condition, valid for $\alpha<1$.

Computing the derivative of Eq.~\eqref{eq:app_null_gradient_condition} leads to
\begin{equation}
    \frac{1}{D_k} \frac{\partial \Bar{u}_k}{\partial h_j} = \frac{1}{N} \sum_{i=1}^N \lambda_{ik} \chi_{ij}\left[h_n\right] \; ,
\end{equation}
where susceptibilities are defined by Eq.~\eqref{eq:susceptibilities_field_dependent}. Merging both expressions, we can obtain a relation between susceptibilities, which taking the limit $\{h_n\}\to0$ can be written as
\begin{equation}
    \chi_{ij} = Y_i \left( \delta_{ij} + 2 \sum_{r=1}^N J_{ir}\chi_{rj} \right) \; ,  \label{eq:relation_susceptibilies} 
\end{equation}
with
\begin{widetext}
\begin{equation}
    Y_i = \frac{s}{2\bar\varepsilon_i^2} \Bigg[  \bar\varepsilon_i B_s(2s\beta\bar\varepsilon_i) + \left(\omega_x + 2\sum_{l=0}^{M-1}\lambda_{il}\bar u_l \right)^2 \left(\frac{s\beta}{2} B_s^\prime (2s\beta\bar\varepsilon_i) - \frac{1}{4\bar\varepsilon_i}B_s(2s\beta\bar\varepsilon_i)\right)  \Bigg] \; .
\end{equation}
\end{widetext}

We can manipulate Eq.~\eqref{eq:relation_susceptibilies} to write a matrix equation,
\begin{equation}
    \sum_{r=1}^N \left( \delta_{ir} - 2Y_iJ_{ir} \right) \chi_{rj} = Y_i \delta_{ij} \; ,
\end{equation}
which allow us to obtain the susceptibilities by inverting the matrix $A_{ij} = \delta_{ij} - 2Y_iJ_{ij}$, writing
\begin{equation}
    \chi_{ij} = A^{-1}_{ij} Y_j \; ,
\end{equation}
as particularized in Eq.~\eqref{eq:susclosedexp} for $\beta\to\infty$.

\section{Classical analysis of the phase transition for $\alpha = 0$}
\label{app:classical}

For $\alpha=0$ the interaction is homogeneous and alternating in sign and we can express Hamiltonian \eqref{eq:Hamiltonian} in terms of total spin operators for each of the sublattices, $J_\Lambda^\gamma = \sum_{i \in \mathcal L(\Lambda)} S_i^\gamma$, with $\Lambda = A, B$ the sublattice index and $\mathcal L(\Lambda)$ the corresponding sublattice,
\begin{equation}
    H = -\omega_z \left(J_A^z + J_B^z\right) -\omega_x \left(J_A^x + J_B^x \right) - \frac{\Gamma}{N}\left(J_A^x - J_B^x\right)^2 \,.
    \label{eq:Hbigspin}
\end{equation}
This Hamiltonian commutes with the total spin operators, $J_\Lambda^2 = (J_\Lambda^x)^2 + (J_\Lambda^y)^2 + (J_\Lambda^z)^2$. Consequently, it connects only states with the same total spins $J_A^2$ and  $J_B^2$. Using exact diagonalization for finite sizes, we show in Appendix \ref{app:exactdiag} that the ground state always lies on the subspace of maximum total spins. This implies that the ground state properties of the model are perfectly captured by a model of two spins of sizes, $j_A = j_B = sN/2$, interacting according to Eq. \eqref{eq:Hbigspin}. In the thermodynamic limit, $N \to \infty$, the description is further simplified by the fact that these spins can be described exactly in their classical limit \cite{lieb1973the}. We can therefore study the energy resulting from replacing the spins with classical magnetizations $J_\Lambda^\gamma \to s N/2 m_\Lambda^\gamma$. The magnetizations are unit vectors that we can assume to lie on the $x, z$ plane $\boldsymbol m_\Lambda = (\cos \theta_\Lambda, 0, \sin \theta_\Lambda)$. With this, the energy per site reads
\begin{equation}
\begin{split}
    e(\theta_A, \theta_B)= & -\frac{s\omega_z}{2} \left(\sin \theta_A + \sin \theta_B\right) \\
    &-\frac{s\omega_x}{2}\left(\cos \theta_A + \cos \theta_B\right) \\
    & - \frac{s^2\Gamma}{4}\left(\cos \theta_A - \cos \theta_B\right)^2 \,.
\end{split} \label{eq:classical_energy}
\end{equation}
The ground state staggered magnetization is given by $m_s = s (\cos \bar \theta_A - \cos \bar \theta_B) / 2$, with $\bar \theta_A$ and $\bar \theta_B$ the minimizers of $E(\theta_A, \theta_B)$.

\begin{figure}
    \centering
    \includegraphics[width=\columnwidth]{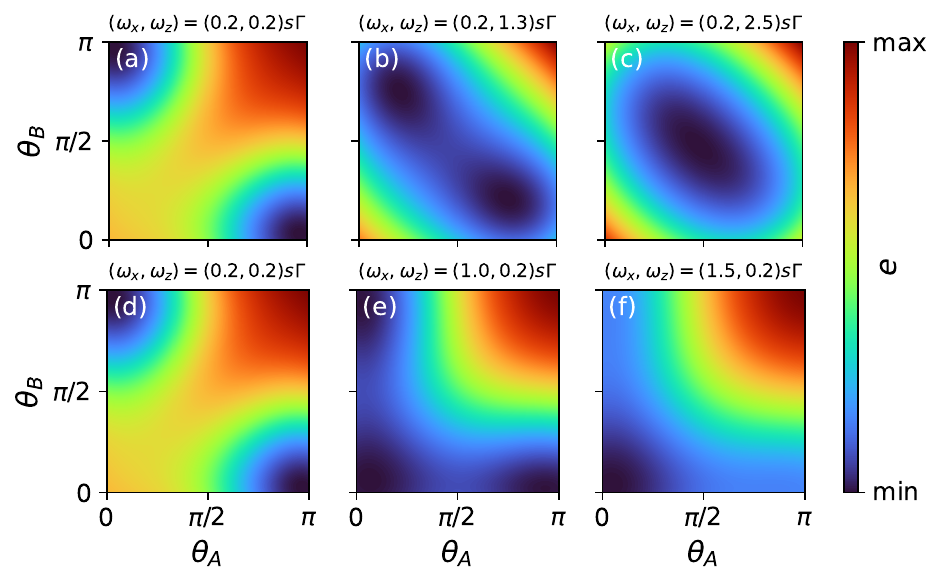}
    \caption{Classical energy landscape [Eq.~\eqref{eq:classical_energy}] for different transverse, $\omega_z$, and longitudinal, $\omega_x$, fields.}
    \label{fig:classical}
\end{figure}

In Fig. \ref{fig:classical} we show the energy landscape as a function of $\theta_A$ and $\theta_B$ for different values of $\omega_x$ and $\omega_z$. For $\omega_x / (s \Gamma) \ll 1$ and $\omega_z / (s \Gamma) \ll 2$ there are two global minima at $(\bar \theta_A, \bar \theta_B) = (0, \pi)$ and $(\bar \theta_A, \bar \theta_B) = (\pi, 0)$ corresponding to the two symmetric antiferromagnetic configurations with $m_s = \pm s$. If $\omega_x$ is kept fixed and $\omega_z$ is increased until $\omega_z / (s \Gamma) > 2$, the two global minima merge into a single one at $(\bar \theta_A, \bar \theta_B) = (\pi/2, \pi/2)$ corresponding to a paramagnetic configuration with $m_s = 0$. The coalescence of minima is indicative of a second-order phase transition. Contrarily, if $\omega_z$ is kept fixed and $\omega_x$ is increased a new local minimum progressively develops at $(\theta_A, \theta_B) = (0, 0)$ until at $\omega_x / (s \Gamma) \geq 1$ it becomes the global minimum, corresponding to a paramagnetic configuration with $m_s = 0$. The former global minima become local minima. The formation of a new local minimum that grows until becoming the global minimum is indicative of a first-order phase transition. In summary, Fig. \ref{fig:classical} shows that the energy landscape encoded in $E(\theta_A, \theta_B)$ displays the typical behavior associated with first- and second-order phase transitions.

We find that the phase diagram predicted by this simple classical model coincides with the phase diagram computed exactly and shown in Fig. \ref{fig:multiplot_phase_diagram}. It is therefore an equivalent formulation of the minimization problem presented in Eq. \eqref{eq:variational_energy}, in terms of two variables $\theta_A, \theta_B$ instead of the single minimization variable $u$. In this sense, we can understand that Eq. \eqref{eq:variational_energy} provides the most concise formulation of the minimization problem, as a univariate function of the order parameter, that lends itself particularly well to the analysis in terms of the Landau theory of phase transitions that we performed in Sec. \ref{sec:phase_diagram}.

\section{Verification using exact diagonalization}
\label{app:exactdiag}

\begin{figure}
    \centering
    \includegraphics[width=\columnwidth]{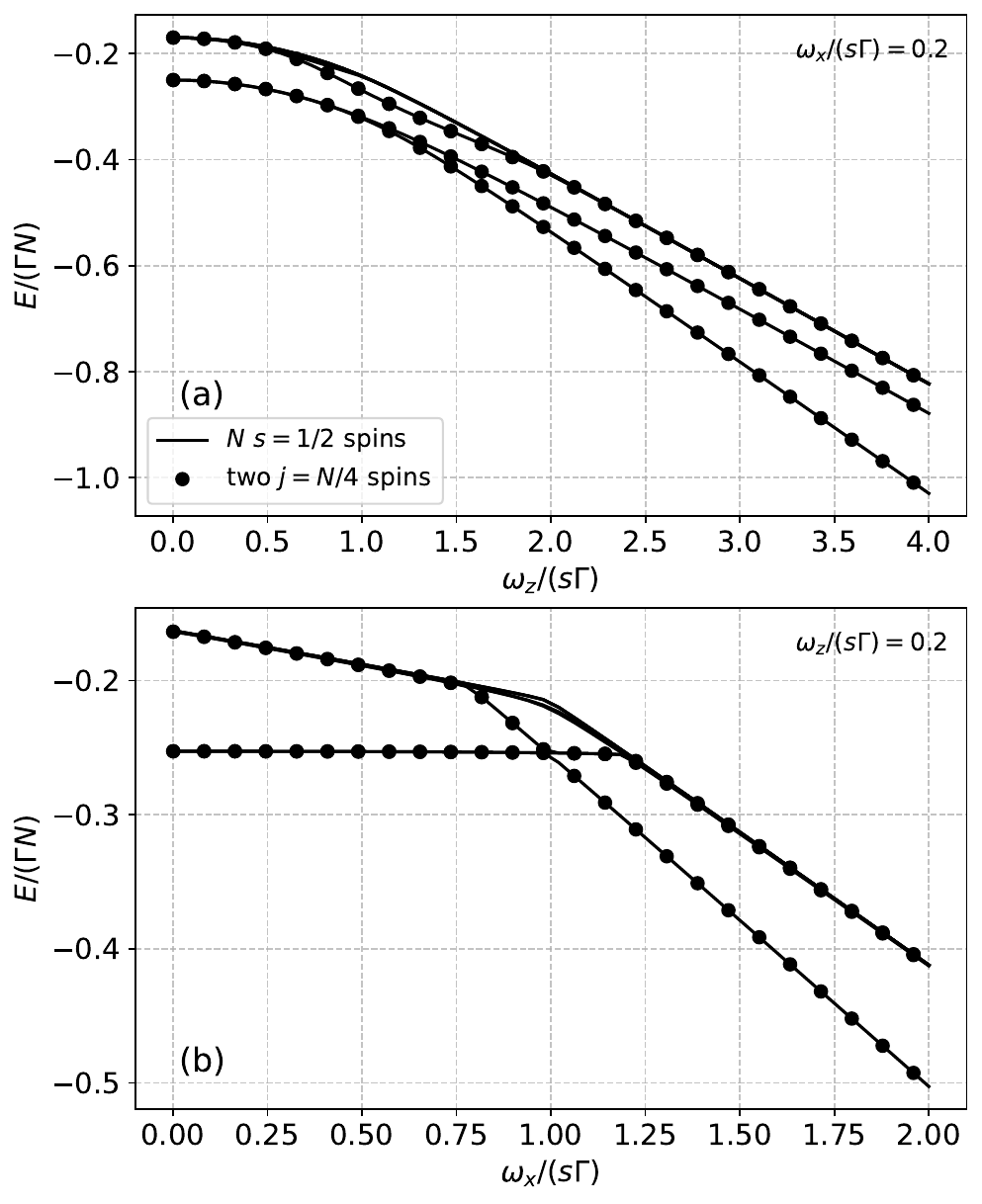}
    \caption{Exact diagonalization spectra of Hamiltonian \eqref{eq:Hamiltonian} for $N = 10$ and $s=1/2$ as a function of $\omega_z$ with fixed $\omega_x/(s\Gamma) = 0.2$ on the top and as a function of $\omega_x$ with fixed $\omega_z/(s\Gamma) = 0.2$ on the bottom. The solid lines correspond to the first 15 energy levels of the full Hamiltonian, expressed as an ensemble of $N$ $s=1/2$ spins. The dots correspond to the first 5 energy levels of the subspace of maximum total spins, expressed as two spins of size $j_A = j_B = sN/2=5/2$ described by Hamiltonian \eqref{eq:Hbigspin}.}
    \label{fig:spectrum_exact_diag}
\end{figure}

For $\alpha = 0$ there are two equivalent descriptions of the model in terms of either individual spins at each site, as in Hamiltonian \eqref{eq:Hamiltonian}, or in terms of total spin operators of each sublattice, as in Hamiltonian \eqref{eq:Hbigspin}. As explained in App. \ref{app:classical}, the total spin of each sublattice, $J_\Lambda^2$, is a conserved quantity and the Hamiltonian is block diagonal with blocks of fixed total spins. Here we diagonalize the subspace of maximum total spins, $j_A = j_B = sN/2$ and compare its low-energy spectrum to the low-energy spectrum of the full Hamiltonian \eqref{eq:Hamiltonian}. In Fig. \ref{fig:spectrum_exact_diag} we show that the ground-state energies of the two Hamiltonians coincide, with differences appearing in the multiplicity of excited states. We show this for $N = 10$ and $s = 1/2$ fixing $\omega_x / (s\Gamma) = 0.2$ and varying $\omega_z / (s \Gamma)$ and vice versa, but the same behavior is observed for any accessible system size, $N$, and spin size, $s$, and for any values of $\omega_x/ (s \Gamma)$ and $\omega_z/ (s \Gamma)$ of the phase diagram. This implies that the low energy physics are well captured by the subspace of maximum total spins, $j_A = j_B = sN/2$. This fact is exploited in App. \ref{app:classical} to provide a classical description of the ground state phase diagram. 

\section{Numerical parameters}\label{app:vithyperparams}
The visual transformer (ViT), the variational ansatz used to obtain the numerical results discussed in the main text, is a type of neural network composed of several blocks. A detailed block diagram with all the operations involved, as well as the hyperparameters that define the architecture, can be found in Ref.~\citenum{rocajerat2024transformer}. For this work the only modification that has been made is to add more attention blocks, the so-called core blocks in said reference, which have been fundamental to capture all the long-range correlations presented by the model. Specifically, we have used three of those blocks. The new training procedure is described in the following:
The optimization process is carried out by Stochastic Gradient Descent (SGD) with custom schedules for the learning rate, $\lambda$, combined with the Stochastic Reconfiguration (SR)  method \cite{sorella2000green}, characterized by a stabilizing parameter called diagonal shift that we denote as $\Delta_{sr}$. The training protocol for $\lambda$ consists of a linear warm-up followed by an exponential decay. This protocol is defined by an initial value of the learning rate, $\lambda_0$, a maximum value, $\lambda_{\mathrm{max}}$ that it attains after $n_{\mathrm{warm}}$ iterations corresponding to the linear warm-up and a ratio $\gamma$ for the exponential decay. To obtain the results presented throughout this manuscript, a maximum of $500$ iterations were used for training, along with the following parameters: $\lambda_0 = 0.1$, $\lambda_{\mathrm{max}} = 6.0$, $n_{\mathrm{warm}} = 150$, $\gamma = 0.999$. The stabilizer parameter present in the SR method was kept constant with the value of $\Delta_{sr} = \num{e-4}$.

\section{Additional numerical results}
\label{app:morenumericalresults}
\begin{figure*}
    \centering
    \includegraphics[width=\textwidth]{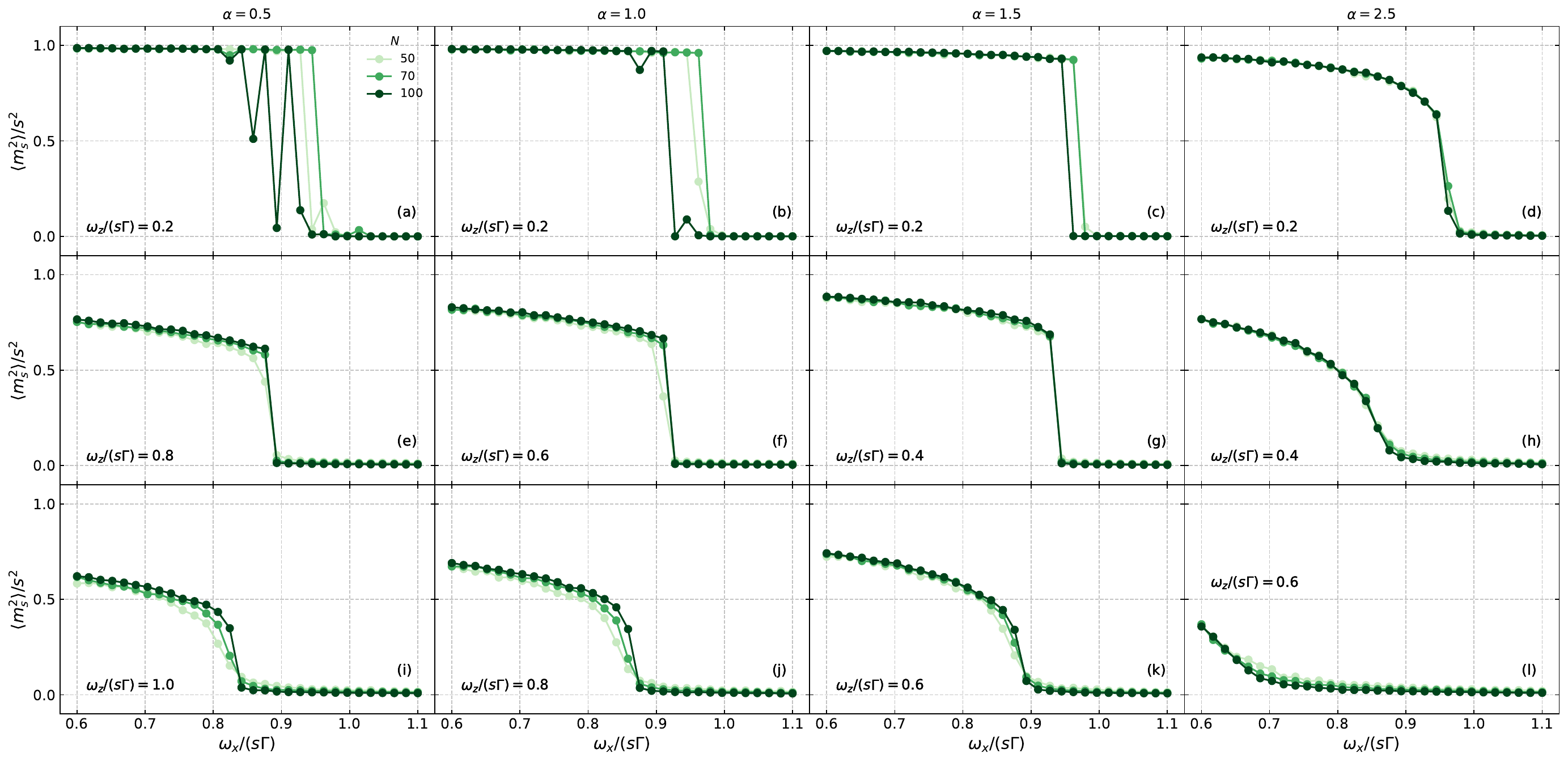}
    \caption{Squared staggered magnetization slices of the phase diagram for different chain sizes. Lighter colors indicate smaller sizes, being $N = 50$ the smallest, $N = 100$ the largest and $N = 70$ the intermediate size.}
    \label{fig:suppnumerical_fig}
\end{figure*}

Figure \ref{fig:suppnumerical_fig} includes further numerical results that reinforce the points discussed in section \ref{sec:numerical} of the main text. There we show different vertical slices of the phase diagram for different values of $\alpha$. These slices allow us to monitor how the tricritical point evolves along the parameter space. These results show that the first-order transition extends to non-negligible transverse fields such as $\omega_z/(s\Gamma) = 0.4$ beyond the strong long-range regime, for $\alpha = 1.5$. The first-order phase transition is present up to $\alpha = 2.5$ and a non-zero transverse field of $\omega_z/(s\Gamma) = 0.2$. In the cases of $\alpha = 0.5$ and $\alpha = 1.0$ we see how the numerical instabilities are alleviated by increasing the transverse field strength.

\bibliography{main.bib}

\end{document}